\documentclass[
aps,
 amsmath,amssymb,
reprint,%
]{revtex4-1}

\usepackage{graphicx}
\usepackage{bm}
\usepackage[normalem]{ulem}
\usepackage{float}
\usepackage{siunitx}

\newcommand{\un}[1]{ \,#1 }
\newcommand{\vect}[1]{ \mathbf{#1} }
\newcommand{\tens}[1]{\mathrm{#1}}

\newcommand{\mean}[1]{\overline{#1}}

\begin{document}

\title[]{Elastic turbulence in two-dimensional Taylor-Couette flows}
\author{R. van Buel}
\email{r.vanbuel@tu-berlin.de}
\affiliation{Technische Universit{\"a}t Berlin - Hardenbergstrasse 36, 10623 Berlin}
\author{C. Schaaf}
\affiliation{Technische Universit{\"a}t Berlin - Hardenbergstrasse 36, 10623 Berlin}
\author{H. Stark}
\affiliation{Technische Universit{\"a}t Berlin - Hardenbergstrasse 36, 10623 Berlin}



\begin{abstract}
We report the onset of elastic turbulence in a two-dimensional Taylor-Couette geometry using numerical solutions of the Oldroyd-B model, also performed at high Weissenberg numbers with the program OpenFOAM. Beyond a critical Weissenberg number, an elastic instability causes a supercritical transition from the laminar Taylor-Couette to a turbulent flow.
The order parameter, the time average of secondary-flow strength, follows the scaling law 
$\Phi \propto (\mathrm{Wi} -\mathrm{Wi}_c)^{\gamma}$ with $\mathrm{Wi}_c=10$ and $\gamma = 0.45$.
The power spectrum of the velocity fluctuations shows a power-law decay with a characteristic exponent, which strongly depends on the radial position.
It is greater than two, which we relate to the dimension of the geometry.
\end{abstract}

\maketitle

\section{Introduction}
Viscoelastic polymer solutions have remarkable properties compared to their Newtonian counterparts. Especially at very small scales, such as employed in microfluidic devices, they enhance mixing \cite{arratia2006elastic, li2010creation,thomases2009transition} and heat transfer \cite{lund2015elastic,zhang2017numerical}. This is due to \textit{elastic turbulence} \cite{groisman2000}, which bears the same qualities as inertial turbulence\cite{groisman2004elastic}. 
The relevant dimensionless number for viscoelastic fluids is the Weissenberg number $\mathrm{Wi}$, which compares the polymer relaxation time to the characteristic time of the flow dynamics \cite{belan2017boundary}.

Elastic turbulence shows many characteristics similar to inertial turbulence \cite{groisman2004elastic}. 
The fluid flow exhibits a significant increase in flow resistance \cite{lumley1973}, random fluctuations of fluid velocity that increase with fluid elasticity, intensified mixing of 
mass, and a broad range of activated temporal or spatial frequencies with power-law scaling\cite{grilli2013transition}.
Berti \emph{et al.} \cite{berti2008two} performed numerical simulations of the Oldroyd-B model in a two-dimensional periodic Kolmogorov flow, at very small Reynolds numbers Re. 
Indeed, above the elastic instability they observed an increase in turbulent drag and a power-law scaling in the velocity spectrum.

The elastic component of the fluid, polymers suspended in a solvent, is affected by two processes: stretching of the polymer molecules by velocity gradients and relaxation of elastic stresses.
The dominant process depends on the value of the Weissenberg number.
At $ \mathrm{Wi} \ll 1$ the relaxation of the polymers is much faster than the stretching time due to velocity gradients.
The polymer acts like a rigid rod and the fluid flow is Newtonian.
For $\mathrm{Wi} > 1$ stretching due to velocity gradients overcomes relaxation. 
Polymers are considerably elongated\cite{lumley1973} and the polymeric stresses grow.
This effect is further enhanced by curved streamlines\cite{pakdel1996elastic,fouxon2003spectra}.

The general consensus is that at low Reynolds numbers the flow of a viscoelastic fluid with straight streamlines is linearly stable but exhibits an elastically induced nonlinear or subcritical instability with increasing Weissenberg number \cite{morozov2005subcritical,pan2013nonlinear,morozov2015introduction}.
This is confirmed by theoretical investigations for a plane Couette flow \cite{morozov2005subcritical,pakdel1996elastic} and in experiments of Pan \emph{et al.} on flow through straight channels \cite{pan2013nonlinear}. The transition is associated with a sudden increase in the velocity fluctuations.

Simulations performed on viscoelastic fluids in a 3D Taylor-Couette geometry but restricted to axisymmetric flows confirmed the occurrence of a standing wave pattern beyond a critical Weissenberg number, as predicted by Shaqfeh\cite{shaqfeh1992}.
The transition to this pattern occurs via a supercritical Hopf bifurcation 
\cite{northey1992finite,avgousti1993non,avgousti1993viscoelastic,avgousti1993spectral}.
However, the most unstable modes occur for the non-axisymmetric case \cite{avgousti1993non,joo1994observations} and 
the bifurcation diagram for both the spiral and ribbon modes were calculated using the Oldroyd-B model \cite{sureshkumar1994non}. 
According to these investigations, the transition of a thin-gap flow to one or both of these modes is always subcritical.
But the authors also showed that for large gaps both modes develop through a linear instability
\cite{sureshkumar1994non}.

In this letter we investigate the occurrence of elastic turbulence of a viscoelastic fluid in a 2D Taylor-Couette geometry using the Oldroyd-B model. By restricting our investigations to two spatial dimensions, we have the simplest implementation of a Taylor-Couette flow.
We work at low Reynolds numbers appropriate for flows at the micron scale. 
Thus, the observed turbulence is only due to the polymeric, \textit{i.e.}, the elastic component of the fluid.
We observe a supercritical transition from the laminar Taylor-Couette to an irregular flow at a critical Weissenberg number $\mathrm{Wi}_c$.
We illustrate the irregular flow and identify it as turbulent by demonstrating a power-law decay of the power spectrum of the velocity fluctuations with an exponent above two.

Our numerical simulations are performed up to high Weissenberg numbers of $\mathrm{Wi} = 200$ using the program OpenFOAM \cite{weller1998tensorial}.
However, numerical simulations at high Wi experience numerical instabilities known as the high-Weissenberg-number problem.  
It is related to the unbounded exponential growth of the conformation tensor which cannot be balanced by convection\cite{fattal2004,fattal2005}.
In particular, regions near stagnation points or strong deformation rates are sensitive to numerical instabilities \cite{fattal2005}.
The simulations can be stabilised by the log-conformation tensor method \cite{afonso2012kernel}, which is implemented via 
the package RheoTool within OpenFOAM, and the choice of an appropriate grid size\cite{fattal2005}. 
This enables stable simulations at high Weissenberg numbers\cite{fattal2004,fattal2005,afonso2012kernel}.

The remainder of the letter is structured as follows. 
First,
our employed methods are presented. We describe the governing equations of the Oldroyd-B model, the geometry, and 
discuss the computer program used for the numerical calculations.
Then,
we present our main results and show the transition to elastic turbulence and the characteristic power spectra of the fluctuating 
velocity field. Finally, we conclude.

\section{Methods}
\begin{figure}[t!]
\begin{center}
        \includegraphics[width=.65\columnwidth]{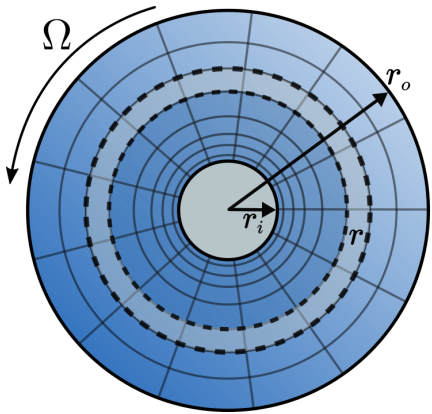}
\end{center}
\caption{Schematic of the 2D Taylor-Couette geometry and the spokes wheel mesh used in OpenFOAM.
We have $N_r = 100$ cells in the radial direction and $N_\theta = 120$ cells in the angular direction.
The radii of the inner and outer cylinders are $r_i = 2.5 \un{\mu m}$ and $r_o = 10 \un{\mu m}$, respectively.
The outer cylinder rotates with an angular velocity $\Omega = 2 \pi \un{{s}^{-1}}$.
For the angular ensemble average, the average of the cells along a specific radius $r$ is taken, which is indicated by a lighter shaded area surrounded by dashed lines.
}
\label{fig:schematic}
\end{figure}

\begin{figure*}[t!]
\centering
\includegraphics[width=.80\textwidth]{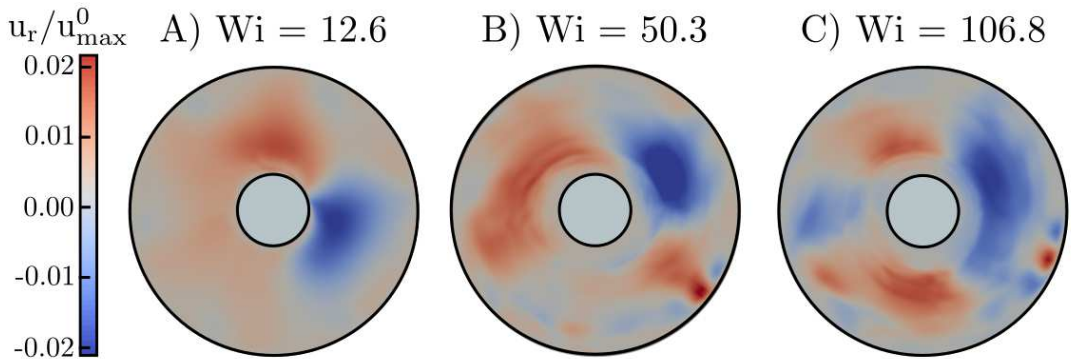}%
\caption{Snapshots of the normalised radial velocity component $u_r / u_\text{max}$ for Weissenberg numbers of A) Wi $ =12.6$, B) Wi $=50.3$ and C) Wi $=106.8$ in the 2D Taylor-Couette geometry and with the maximum velocity $u_\text{max} = \Omega r_\mathrm{o} = 2 \pi \cdot {10 \un{\mu m /s}}$.
}
\label{fig:snapshots}
\end{figure*}


We consider a viscoelastic fluid in a 2D Taylor-Couette geometry consisting of two concentric cylinders with $r_o$ the radius of the outer cylinder and $r_i$ the radius of the inner cylinder. The outer cylinder is rotated counter-clockwise with a constant angular velocity $\Omega$ and the inner cylinder is fixed. A schematic of our set-up can be seen in fig.~\ref{fig:schematic}.
The dynamics of the flow field $\vect{u}(\vect{r},t)$ as a function of the position $\vect{r}$ and time $t$ can be described with a generalised Navier-Stokes equation already formulated with unitless quantities:
\begin{equation}
\label{eq:NS}
\mathrm{Re}\left( 
\frac{\partial \vect{u}}{\partial t} + \left(\vect{u} \cdot \nabla\right) \vect{u} \right)
= - \nabla p + \nabla^2 \vect{u} +  {\nabla \bm{\tau}} \, .
\end{equation}
Here, $p$ is the dimensionless pressure
and we expressed lengths in units of $r_o$, times in units of $\Omega^{-1}$, and velocities in units of $r_o \Omega$ so that the Reynolds number becomes \mbox{$\mathrm{Re} = \rho \Omega r_o^2 / \eta_s$},
where $\rho$ is density and $\eta_s$ the shear viscosity of the solvent.
The additional viscoelastic stresses due to the dissolved polymers are described by the polymeric stress tensor $\bm{\tau}$,
for which we choose  the constitutive relation of the linear Oldroyd-B model written in dimensionless form \cite{oldroyd1958non,oldroyd1950formulation}:
\begin{equation}
\label{eq:Oldroyd}
\bm{\tau} + \mathrm{Wi} \overset{\nabla}{\bm{\tau}} = 
{
\beta \left[ \nabla \otimes \vect{u} + (\nabla \otimes \vect{u})^\mathrm{T} \right] \, ,
}
\end{equation}
where $\overset{\nabla}{\bm{\tau}}$ is the upper convective derivative of the 
stress tensor
defined as
\begin{equation}
\overset{\nabla}{\bm{\tau}} = \frac{\partial  \bm{\tau}}{\partial t} + \vect{u}\cdot \nabla\bm{\tau} 
- (\nabla \otimes \vect{u})^\mathrm{T} \bm{\tau} - \bm{\tau}  ( \nabla \otimes \vect{u}) \, .
\end{equation}
The quantity $\beta = \eta_p/\eta_s$ denotes the ratio of the polymeric shear viscosity to the solvent shear viscosity.
Furthermore, we introduce the Weissenberg number $\mathrm{Wi} =  \lambda \dot{\gamma}$, with $\lambda$ the characteristic relaxation time of the dissolved polymers and $\dot{\gamma}$ the shear rate.
For the Taylor-Couette geometry the characteristic shear rate is the angular velocity of the outer cylinder 
$ \dot{\gamma} = \Omega$ and we have $\mathrm{Wi} = \lambda \Omega$.

Considering an axisymmetric flow, which corresponds to our 2D geometry, an analytic solution of eqs.~(\ref{eq:NS}) and (\ref{eq:Oldroyd})
has been found\cite{larson1990purely}.
It agrees with the Taylor-Couette flow of a Newtonian fluid at low Re in the same geometry.
The solution, which we denote as the base flow $\vect{u}^0$, is given by the simple shearing flow
\begin{equation}
 u_r^0 = 0  \, ; \qquad u_\phi^0 = A r + B r^{-1} \, ,
\label{eq.TCflow}
\end{equation}
with
\begin{equation}
 A = \frac{r^2_o }{r^2_o-r^2_i}\Omega  \, ; \qquad B = -\frac{r^2_i r^2_o}{r^2_o-r^2_i}\Omega \, .
\end{equation}
Moreover, for the Oldroyd-B model the components of the polymeric stress tensor for the base flow are given by
\begin{align}
&\tau_{rr}^0 = 0 \, , \\
&\tau_{r\phi}^0 = -2 \eta_p B r^{-2} \, , \\
&\tau_{\phi\phi}^0 = 8 \eta_p \lambda B^2 r^{-4} \, .
\end{align}
For later use we write them in real units.
The base solution becomes unstable for Weissenberg numbers above a critical Weissenberg number $\mathrm{Wi}_c$.

\subsection{OpenFOAM}
In our work, eqs.~(\ref{eq:NS}) and (\ref{eq:Oldroyd})
are solved using the program OpenFOAM, which is an open-source finite-volume solver for computational fluid dynamics simulations on polyhedral grids. We adopt a specialised solver for viscoelastic flows implemented in OpenFOAM.
It is called rheoTool, which is based on a solver for complex fluids develop by Favero et al. \cite{favero2010}. The rheoTool solver has been tested for accuracy in benchmark flows and it has been shown to have second-order accuracy in space and time \cite{pimenta2017}.

However, simulations of viscoelastic fluid simulations experience stability issues, especially at high Weissenberg numbers, where the conformation tensor eventually loses its positive definiteness due to numerical discretisation errors.
To tackle this problem, the log-conformation tensor approach \cite{fattal2004} is implemented in the rheoTool solver. 
This approach employs a change of variables for the conformation tensor $\tens{A}$, which is a positive definite tensor and related to the polymeric stress tensor by 
\begin{equation}
\label{eq:afine}
\bm{\tau} = \frac{\beta}{\mathrm{Wi}} (\tens{A} - \tens{I}) \, .
\end{equation}

Now, the main idea consists in introducing the (tensor) logarithm of $\tens {A}$, $\tens{\Theta} = \ln \tens{A}$,
which does not have to be positive definite, and in transforming the constitutive relation (\ref{eq:Oldroyd}) in order to obtain a dynamic
equation for $\tens{\Theta}$. To do so, one splits the transpose of the velocity gradient tensor into a pure deformational 
component (the symmetric tensor $\tens{B}$, which necessarily commutes with $\tens{A}$), a pure rotational component
(the antisymmetric tensor $\tens{\Omega}$), and an additional term:
\begin{equation}
(\nabla \otimes \vect{u})^{\mathrm{T}} = \tens{B} + \tens{\Omega} + \tens{N} \tens{A}^{-1} \, .
\end{equation}
The last term on the right-hand side means that $\tens{B}$ and $\tens{\Omega}$ are not the usual deformation and vorticity tensors associated with $\nabla \otimes \vect{u}$. According to Ref.~\cite{fattal2004} the decomposition always 
exists and is unique. Using this decomposition, eq.~(\ref{eq:Oldroyd}) transforms into a constitutive relation for $\tens{\Theta}$,
\begin{equation}
\label{eq:logconst}
\frac{\partial\tens{\Theta}}{\partial t} + \vect{u}\cdot \nabla\tens{\Theta} 
-(\tens{\Omega} \tens{\Theta}  - \tens{\Theta} \tens{\Omega}) - 2 \tens{B}
= \frac{1}{\mathrm{Wi}} (\mathrm{e}^{-\tens{\Theta}} - \tens{I}) \, .
\end{equation}
More details of the method are given in Ref.~\cite{fattal2004}. 
Evolving $\tens{\Theta}$ in time and then transforming back to the conformation and stress tensors leads to generally enhanced stability \cite{pimenta2017} and, in particular, minimizes stability issues for flows at high Weissenberg numbers \cite{fattal2004,fattal2005}. 
However, this comes at a cost of exponentially increasing errors in solving eq.~(\ref{eq:logconst}) and therefore requires a sufficiently fine grid, which we checked in our simulations. 
Figure\ \ref{fig:schematic} shows the spokes wheel grid used in OpenFOAM with $N_r = 100$ cells in the radial direction and $N_\theta = 120$ cells in the angular direction. We use a mesh refinement towards the inner cylinder, where velocity gradients become larger.

\subsection{Simulation parameters}
We give all the simulation parameters in real mks units as required by OpenFOAM.
To ensure that the observed turbulent behaviour is due to the polymeric contribution, we choose
our simulation set-up such that fluid inertia is negligible ($\mathrm{Re}\ll 1$). The outer radius is $r_o=\SI{10}{\mu m}$, the inner radius is $r_i = \SI{0.25}{r_o}$, 
and together with a rotational frequency of ${\Omega=\un{2\pi/s}}$ this leads to a small Reynolds number $\mathrm{Re} \approx 10^{-4}$. 
The ratio of the polymeric and solvent viscosities is set to $\beta={\eta_p}/{\eta_s}=1.5$. 
Furthermore, we adjust the Weissenberg number by varying the polymeric relaxation time $\lambda$ between \SI{0.4}{s} and \SI{40}{s}, which corresponds to Weissenberg numbers between \un{2.5} and \un{250}. 
The thickness of the grid cells in the radial direction $\Delta r$ is ${2.0}\un{\cdot 10^{-3}}\, r_o$ at the inner cylinder and increases to ${1.9}\un{\cdot 10^{-2}}\,r_o$ at the outer cylinder. The time step of the simulation is $\delta t = 10^{-5}\un{s}$, where the velocity, pressure, and stress fields are extracted every 5000 steps. The simulation time ran at least \SI{100}{s} and up to \SI{600}{s}, with an average of \SI{250}{s}.

\section{Results}
Our numerical simulations show that for increasing fluid elasticity, \textit{i.e.}, increasing Weissenberg number Wi,
the Taylor-Couette or base flow becomes unstable. At low Wi numbers we recover the simple shear flow corresponding to the analytic result presented in eq.~(\ref{eq.TCflow}).
Upon increasing the polymeric relaxation time $\lambda$, a secondary flow is created, which strongly fluctuates in time and which has a non-zero radial component. 
Snapshots of the normalised radial velocity component $u_r / u_\text{max}$ are presented in fig.~\ref{fig:snapshots} for three different values of the Wi number above the critical number, where the instability occurs:
A) $\mathrm{Wi}=12.6$ (right above the transition),
B) $\mathrm{Wi}=50.3$ and C) $\mathrm{Wi}=106.8$. 
The snapshots show that the radial symmetry of the base flow is broken and reveal the existence of small vortices for the cases B and C visible as consecutive blue and red circular patches.
Videos M1 and M2 of the Supplemental Material 
show the time evolution of the radial and azimuthal velocity fields for case C.

\subsection{Transition to elastic turbulence}

\begin{figure}[t!]
\centering
  \includegraphics[width=.42\textwidth]{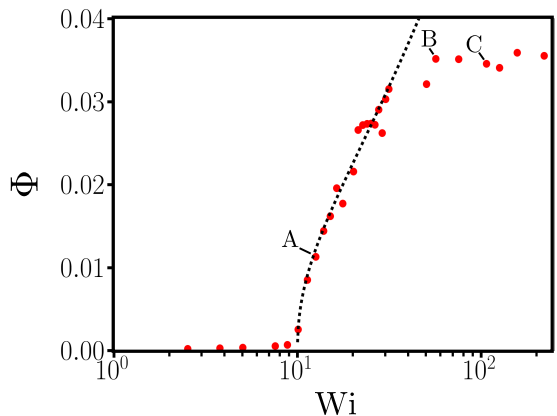}%
\caption{Elastic instability:
Order parameter $\Phi =\mean{\sigma_{r,\theta}}$ as a function of the Weissenberg number Wi, with the secondary-flow strength $\sigma_{r,\theta}$ 
defined in eq.~ (\ref{eq.standard}).
The dashed line shows the scaling law $(\mathrm{Wi} -\mathrm{Wi}_c)^{\gamma}$ with $\mathrm{Wi}_c=10$ and $\gamma = 0.45$.
The positions A, B, and C correspond to the snapshots of fig.~\ \ref{fig:snapshots}.
}
\label{fig:orderparameter}
\end{figure}

To investigate the transition, we define an order parameter 
$
\Phi = \mean{\sigma_{r,\theta}},
$
which is a time average of the secondary-flow strength
\begin{equation}
\sigma_{r,\theta}(t) \equiv\left. \sqrt{\left\langle [\vect{u}(\vect{r},t)-\vect{u^0}(\vect{r})]^2 \right\rangle_{r,\theta}}\right/u^0_{\mathrm{max}} \, ,
\label{eq.standard}
\end{equation}
where we take the square root of the second moment of the flow field about the base flow $\vect{u^0}(\vect{r})$.
Here, the subscript $\vect{x}$ in $\langle \dots \rangle_{\vect{x}}$ denotes the coordinates over which the average is taken and $u^0_{\mathrm{max}}$ is the velocity of the outer cylinder. The order parameter is plotted in fig.~\ref{fig:orderparameter} as a function of the Weissenberg number.
A clear transition from the laminar base flow ($\Phi=0$) to the occurrence of a secondary flow ($\Phi>0$) is observed upon increasing the elasticity of the fluid beyond a critical value of $\mathrm{Wi}_c = 9.99 \pm 0.08$.
The transition {sets in with} a sharp increase in the order parameter that scales as 
$(\mathrm{Wi} -\mathrm{Wi}_c)^{\gamma}$, where $\gamma = 0.45 \pm 0.03$.
It is continuous and therefore supercritical. For clarity, the Weissenberg numbers related to the snapshots of fig.~\ref{fig:snapshots} are indicated with their corresponding letters.

\begin{figure}[t!]
    \centering
        \includegraphics[width=.42\textwidth]{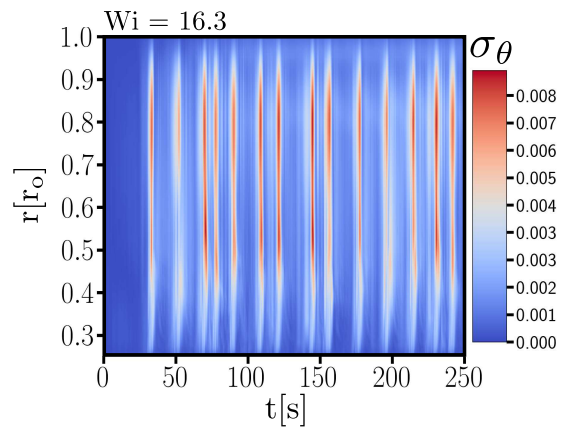}
    \caption{Colour-coded space-time plot of the secondary-flow strength $\sigma_\theta (t)$
    versus the normalised radius $r/r_o$ and time $t$ for $\mathrm{Wi}= 16.3$.
}
\label{fig:velocityfluc_2D}
\end{figure}

We now study the flow field beyond the elastic instability in more detail.
Considering the radial symmetry of our geometry, we investigate the secondary-flow strength of eq.~(\ref{eq.standard}), but only perform an angular average.
The resulting $\sigma_\theta (t)$ is plotted in fig.~\ref{fig:velocityfluc_2D} as a function of the normalised radius $r/r_o$ and time $t$.
The colour-coded $\sigma_\theta (t)$ shows bursts of velocity  in time, which mostly occur simultaneously across the entire radius of the cylinder. 
The intensity varies with the radial location and has broad maxima at around $r\approx 0.55\, r_o$ and $r\approx 0.8 \, r_o$.

\begin{figure*}
    \centering
        \includegraphics[width=.8\textwidth]{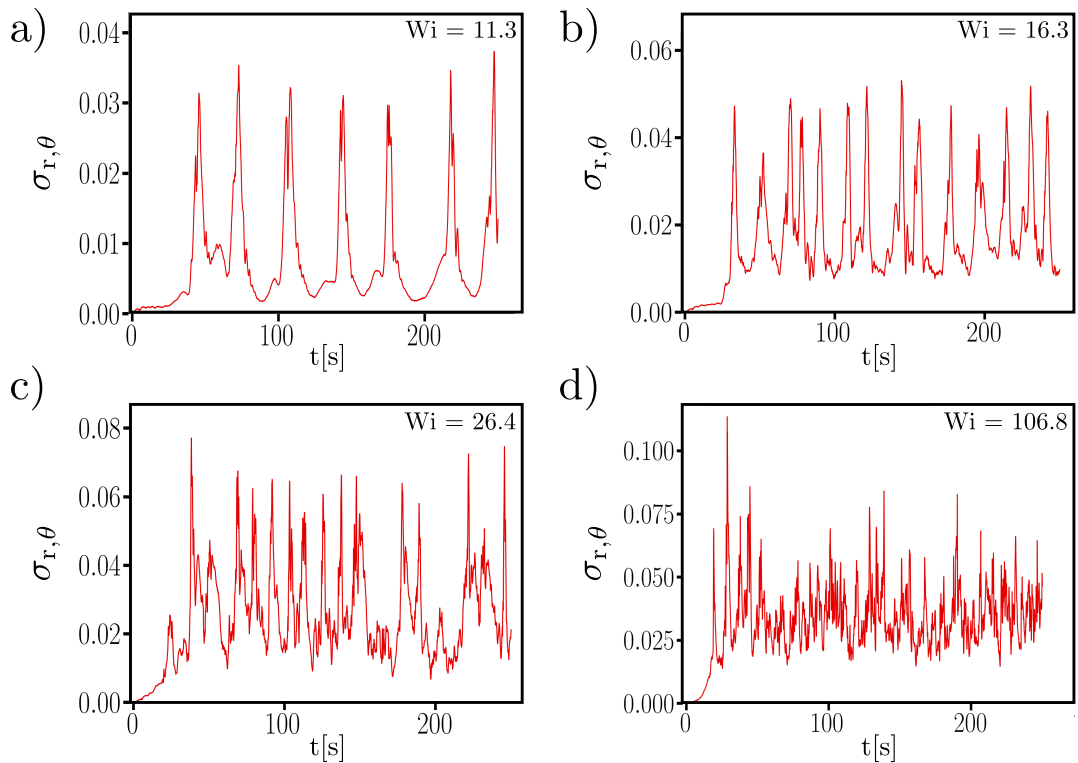}  
\caption{ 
Full cylinder averaged secondary-flow strength $\sigma_{r,\theta} (t)$ plotted as a function of time $t$ for four values of the Weissenberg number: a)~$\mathrm{Wi} = 11.3$, b)~$\mathrm{Wi}= 16.3$, c)~$\mathrm{Wi} = 26.4$ and d)~$\mathrm{Wi} = 106.8$. 
}
\label{fig:velocityfluc_WI}
\end{figure*}

\begin{figure}
\centering
  \includegraphics[width=.35\textwidth]{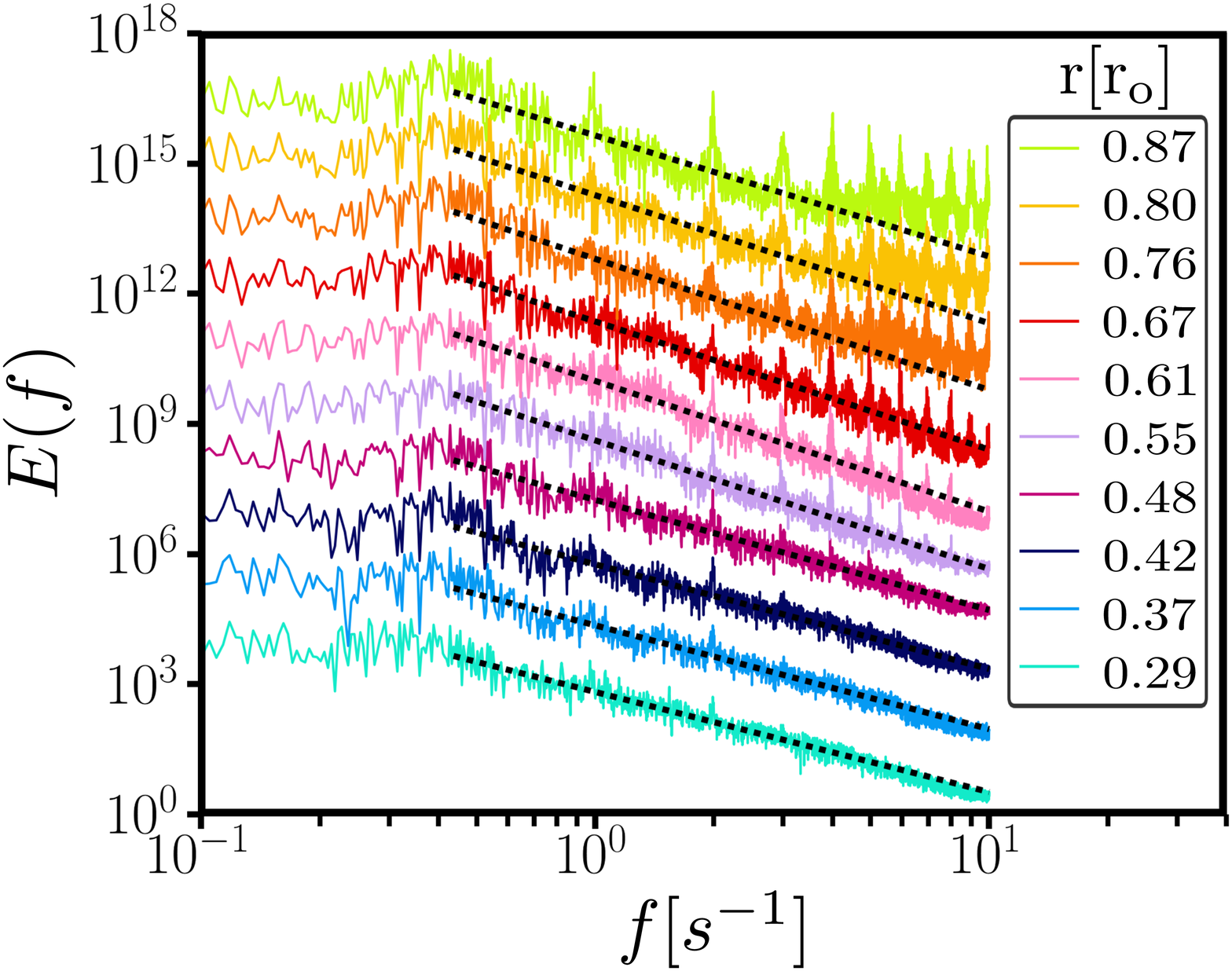}
\caption{
  Power 
  {spectral} 
  density $E(f) = \langle|\Delta\vect{u}(f)|^2 \rangle_\theta$ of the velocity fluctuations $\Delta\vect{u}(r,t)$, ensemble averaged over the 
  {azi\-mu\-thal}
  angle $\theta$, as a function of the frequency $f$.
  {Spectral densities for different radial positions $r$ are shown, all at $\mathrm{Wi} = 25$.}
  The data is arbitrarily offset by multiplying with a constant for clarity.
  {Power laws $f^{-\alpha}$ are fitted and plotted as black dotted lines
  in order to determine the negative slopes $\alpha$.}
}
  \label{fig:PSD}
\end{figure}

\begin{figure}
\centering
  \includegraphics[width=.45\textwidth]{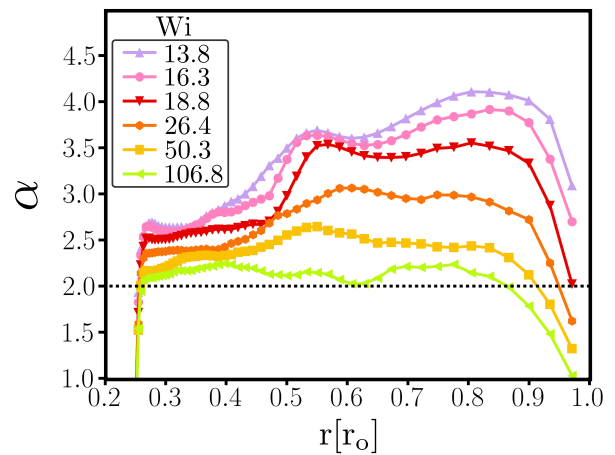}
\caption{
The power-law exponent $\alpha$ of the power spectral density of the velocity fluctuations plotted as a function of the normalised radius $r/r_\text{o}$ for several values of the Weissenberg number
$\mathrm{Wi}$ above $\mathrm{Wi}_c = 10$.}
\label{fig:exponent}
\end{figure}

Increasing the Weissenberg number the secondary flow becomes stronger and more irregular. To illustrate the development of an irregular flow pattern, we plot the fully averaged secondary-flow strength $\sigma_{r,\theta}(t)$ as a function of time in fig.~\ref{fig:velocityfluc_WI}.
The velocity bursts are clearly visible for all Weissenberg numbers. At Weissenberg numbers close to the elastic instability the bursts are nearly regular (a). With increasing Wi they become more  irregular, more frequent and stronger (b), (c) until the secondary flow looks fully turbulent (d).

\subsection{Power spectral density of velocity field}

A determining feature of elastic turbulence is the power-law dependence of the power spectral density of the velocity fluctuations, $E(f) \sim f^{-\alpha}$, with $\alpha > 3$ \cite{groisman2000}.
In fig.~\ref{fig:PSD} the spectral density $E({f}) = \langle | \Delta\vect{u}(f)|^2 \rangle_\theta$, where $\Delta\vect{u}(f)$ is the Fourier transform of $\Delta\vect{u}(\vect{r},t) = \vect{u}(\vect{r},t) - \vect{u^0}(\vect{r})$, is plotted as a function of the frequency $f$.
Here, $\langle \ldots \rangle_\theta$ again indicates the angular ensemble average and $\mathrm{Wi}=25$.
We observe a plateau value in the power spectrum for low frequencies followed by a power-law decay at high frequencies. We attribute the characteristic peaks of $E(f)$ near the outer cylinder to the characteristic frequencies $f_n = n \Omega / 2\pi = n \un{s}^{-1}$ associated with the rotation of the outer cylinder and its higher harmonics.

We observe that the power-law exponent $\alpha$ depends on the radial position and shows non-monotonic behaviour. The radial dependence of $\alpha$ is plotted in fig.~\ref{fig:exponent} for several Weissenberg numbers $\mathrm{Wi} > \mathrm{Wi}_c$. 
Two bulk regions with different exponents $\alpha$ can be distinguished at low Wi, one close to the inner cylinder for radii $r_i < r < 0.55 \, r_o$ and one close to the outer cylinder for radii $0.55\, r_o < r < \, r_o$.
Upon increasing Wi the value of the exponent decreases for both regions and becomes nearly constant for $\mathrm{Wi} > 100$.

The characteristic exponent $\alpha$ of elastic turbulence has been shown to obey the condition $\alpha > 3$. Ref.\cite{fouxon2003spectra} provides proof of the inequality and in experiments values of $\alpha = 3.3-3.5$ \cite{groisman2000,groisman2001stretching} and $\alpha = 3-4$ \cite{sousa2018purely} have been observed.
However, these values correspond to 3D geometries. Based on similar arguments as in Ref.~\cite{fouxon2003spectra} the lower bound of the exponent should be related to the dimension $d$ of the system such that $\alpha > d $.
This is nicely confirmed by our observations, where the exponent is always larger than 2 in agreement with our 2D geometry.

\section{Conclusion}

In this letter we have investigated the onset of elastic turbulence of a viscoelastic fluid in a two-dimensional Taylor-Couette flow by numerically solving the Oldroyd-B model with the program OpenFOAM up to very high Weissenberg numbers.
A supercritical transition from a stable laminar flow to a chaotic flow occurs above a critical Weissenberg number $\mathrm{Wi_c} = 10.0$. The transition is indicated by an order parameter, the time-averaged secondary-flow strength, which follows the scaling law $\Phi \propto (\mathrm{Wi} -\mathrm{Wi}_c)^{\gamma}$ with $\gamma = 0.45$.
The Weissenberg number is increased by increasing the polymeric relaxation time $\lambda$ and the secondary turbulent flow has a non-zero radial component, which is ideal for mixing at the micron scale.

We have further observed the characteristic power-law dependence $E(f) \sim f^{-\alpha}$ in the power spectra of the fluctuating velocity fields, which exhibit irregular bursts in time.
The exponent $\alpha$ shows non-monotonic behaviour as a function of the radial position and we observe two bulk regions 
with different characteristic exponents,
which merge into a single bulk value for the fully turbulent flow upon increasing $\mathrm{Wi}$.
In contrast to 3D elastic turbulence, where the characteristic exponent $\alpha$ obeys $\alpha > 3$, we observe $\alpha > 2$, which we attribute to the smaller spatial dimension of our set-up following the argumentation of Ref.~\cite{fouxon2003spectra}.

Our work clearly demonstrates and quantifies the onset and development of elastic turbulence in an exemplary two-dimensional set-up.
It might be used as a reference for future studies. It also guides the path for using the program OpenFOAM for numerical investigations of elastic turbulence in different geometries and for exploring features such as enhanced turbulent drag or turbulent mixing.

\begin{acknowledgments}
We thank A. Morozov for encouraging discussions and acknowledge support from the Deutsche Forschungsgemeinschaft in the framework of the Collaborative Research Center SFB 910.
\end{acknowledgments}

\bibliography{literature}

\end{document}